\documentclass{aa}
\usepackage{graphics,epsfig}
\usepackage{amssymb}

\newcommand{\kms}{km~s$^{-1}$}

\newcommand{\sgr}{Sgr~B2}

\newcommand{\acet}{CH$_3$CHO}

\begin{document}
\title{Widespread acetaldehyde near the Galactic Centre}
\titlerunning{Widespread acetaldehyde near the Galactic Centre}
\author{Jayaram N Chengalur\inst{1}\thanks{chengalu@ncra.tifr.res.in},
Nissim Kanekar \inst{2}\thanks{nissim@astro.rug.nl}}
\authorrunning{Chengalur \& Kanekar}
\institute{National Centre for Radio Astrophysics, Post Bag 3, Ganeshkhind, Pune 411 007 
\and Kapteyn Institute, University of Groningen, Post Bag 800, 9700 AV Groningen }
\date{Received mmddyy/ accepted mmddyy}
\offprints{Jayaram N Chengalur}
\maketitle
\begin{abstract}  We present Giant Meterwave Radio Telescope images of the 
1065~MHz emission from the $1_{11}\rightarrow$ 1$_{10}$ rotational transition
of acetaldehyde (\acet) in the molecular cloud complex \sgr. Our
observations are unique in that they have a high spatial resolution 
($\sim 4^{''}$), while still being sensitive to large-scale emission. Most 
complex organic molecules in this cloud (e.g. acetone, methyl formate, 
acetic acid) are concentrated in a very small core, $\sim 0.1$~pc across.
In contrast, acetaldehyde is found to be spread over a region at least 100
times larger in extent.  The line emission is confined to regions with 
radio continuum emission and correlates well (in both position and
velocity) with formaldehyde absorption towards this continuum; this is 
consistent with earlier single dish results suggesting that it is likely to be 
weakly mased. Our observations also suggest that grain mantle destruction
by shocks plays an important role in the observed gas phase abundance of \acet~in \sgr.
\keywords{Galaxy: centre --
          ISM: clouds --
          ISM: molecules --
          Astrochemistry --
          radio lines: ISM}
\end{abstract}

\section{Introduction}
	
	The giant molecular cloud complex \sgr~is one of the most interesting 
regions in the Galaxy. Located close to the Galactic Centre, the complex is 
undergoing extensive star formation and hosts a myriad of large molecules, 
many of which have not been detected elsewhere in the Galaxy. 
Most of the complex organic  molecules found here are confined to a compact 
(size $\sim 0.1$~pc) region called the Large Molecular Heimat (LMH) 
(e.g. \cite{miao95,mehringer97a,snyder02}). This is in keeping 
with models of the formation of complex molecules, according to which these
molecules are either directly formed on the surfaces of dust grains or are 
formed from seed molecules which are themselves formed on grain surfaces 
(e.g. \cite{charnley92, charnley95}). When these dust grains are heated 
(typically by radiation from hot stars that form in molecular cloud cores), 
the grain mantles evaporate, releasing the molecules in gaseous form. 
Further, in time dependent astro-chemical models, these molecules are 
rapidly destroyed either by photodissociation or by collisions with other 
molecules. As a result, complex organic molecules are generally found only
in hot molecular cloud cores.

	Recently, however, there have been indications that some large 
organic molecules are more wide-spread in the \sgr~ cloud than was hitherto 
believed. A comparison of the measured flux density from single dish and 
interferometric observations of the millimetre wavelength emission from 
glycolaldehyde suggests that the molecule is distributed over a region at least  
$\sim 1$~pc in size (\cite{hollis01}). However, no direct images of its 
spatial distribution are available. Similarly, emission studies of ethanol
towards selected regions around \sgr~ indicate that it too is likely 
to be wide-spread in this region, but again, no direct images are available 
for this molecule (\cite{mp01}). 

We present here wide field Giant Meterwave Radio Telescope
(GMRT) images of the 1065~MHz $1_{11}\rightarrow$ 1$_{10}$~K transition of 
acetaldehyde (\cite{kleiner96}) towards \sgr; this transition was originally
detected towards \sgr~by Gottleib (1973), using a single dish telescope. 
Acetaldehyde is one of the molecules whose interstellar production mechanism 
is presently quite unclear, placing strong constraints on models of interstellar 
chemistry. It has been found so far in both cold dust clouds as well as 
in hot cores (\cite{ikeda01,matthews85}). Towards the Galactic Centre, 
it has been detected at mm wavelengths in \sgr(N) (\cite{ikeda01}), along 
with its isomers vinyl alcohol and ethylene oxide (\cite{dickens97,turner01}). 
Interferometric observations of the 1065~MHz transition were not possible until 
very recently, as there were no telescopes equipped with receivers at this frequency. 
The newly comissioned GMRT does operate at this frequency, however, and, unlike 
millimetre-wave interferometers, provides both a (relatively) large field 
of view as well as high spatial resolution. Our images show direct evidence 
for the presence of acetaldehyde over a large region in \sgr.  

\section{Observations and Data Analysis}
\label{sec:obs}

	The GMRT (\cite{swarup91}) observations were carried out on 12 May 2001, 
during the commissioning phase of the telescope. The array has a hybrid 
configuration with 14 of its 30 antennas located in a central compact array 
with size $\approx$ 1 km ($\approx$ 3.5 k$\lambda$ at 1065~MHz) and the remaining
antennas distributed in a roughly ``Y'' shaped configuration, giving a 
maximum baseline length of $\approx$ 25~km ($\approx$ 90 k$\lambda$
at 1065~MHz). The baselines between antennas in the central array are similar 
in length to those of the ``D'' array of the VLA while the baselines 
between the arm antennas are comparable to those of the VLA ``B'' array.
An observing bandwidth of 1~MHz ($\sim 280$~km/s) was used for the \acet~
observations, centered at 1065.075~MHz, the line rest frequency 
(\cite{gottlieb73,kleiner96}). The band was divided into 128 spectral 
channels, giving a channel spacing of $\sim 2.2$~\kms. The standard calibrators
3C286 and 3C48 were observed at the start and end of the run respectively in order
to set the absolute flux density scale and also to determine the bandpass shape,
while the compact source 1751$-$253 was used for phase calibration. The total 
on-source time was $\sim 4.2$~hours. In a later, short observing run, the 
system temperature was measured towards the calibrators as well as the source 
by firing noise diodes.  

	The data were converted from the raw telescope format into FITS and
then analyzed in AIPS. After editing the obviously bad data, the 
flux density scale and instrumental phase were calibrated using standard AIPS 
procedures.  Care was also taken to account for the fact that the system 
temperature in the direction of \sgr~ is higher than that towards the flux 
or phase calibrators (due to the wide-spread continuum emission from the 
environs of \sgr~and its proximity to the Galactic Centre). After the initial
calibration, a continuum image was made using the line-free channels and 
used to self-calibrate the U-V visibilities. This was carried out in an 
iterative manner until the quality of the image was found to not improve
on further self-calibration. The continuum emission was then subtracted from the
multi-channel U-V data set and the continuum-subtracted data mapped in all
channels; any residual continuum was then subtracted in the image plane by
fitting a linear baseline to line-free regions. 

	The hybrid configuration of the GMRT array implies that a single GMRT 
observation yields information on both small and large angular scales. Image 
cubes were therefore made at various (U,V) ranges and analysed separately.
The low resolution images (such as Fig.~\ref{fig:uv10ov}) were deconvolved using 
the standard CLEAN deconvolution algorithm to produce the final spectral cube. 
However, at the highest resolutions (such as that of Fig.~\ref{fig:uv60m0}), 
the signal to noise ratio (SNR) is too low and the emission too diffuse for the 
CLEAN algorithm to work reliably; this image has hence not been deconvolved. 
Despite this, the low SNR of this image implies that the inability to 
deconvolve it does not greatly degrade its dynamic range or fidelity; the 
morphology of the emission should hence be accurately traced, apart from an 
uncertainty in the scaling factor (this essentially arises because the main 
effect of deconvolving weak emission at about the noise level corresponds to 
multiplying by a scale factor; \cite{jorsater95,rupen99}).

\section{Results and Discussion}
\label{sec:res}

\begin{figure}[t!]
\begin{center}
\epsfig{file=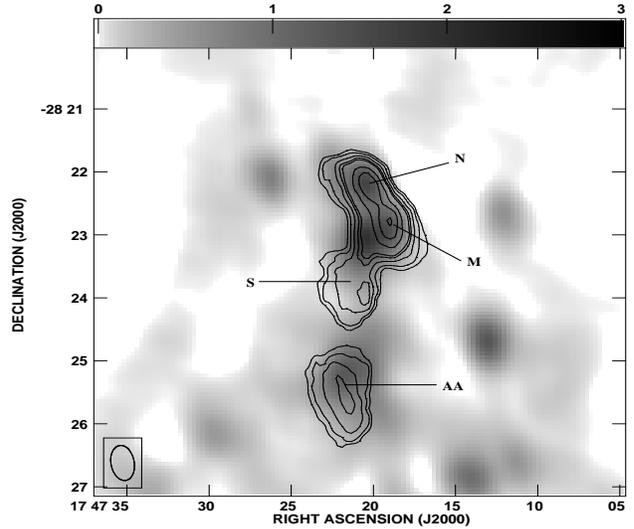,height=3.3truein,width=3.3truein}
\end{center}
\vskip -0.2 in
\caption{Integrated \acet~ emission towards \sgr~ (contours) overlaid on
the 1065~MHz continuum emission (greyscale). The spatial resolution is $34'' 
\times 19''$. The greyscale ranges from 0 to 3.0~Jy/Beam. The contours are at
90, 150, 250, 350, 415, 710, 1000 and 1275~Jy/Beam~m/s. Spectra 
towards the four marked regions are shown in Fig.~2}
\label{fig:uv10ov}
\end{figure}

\begin{figure}[t!]
\begin{center}
\epsfig{file=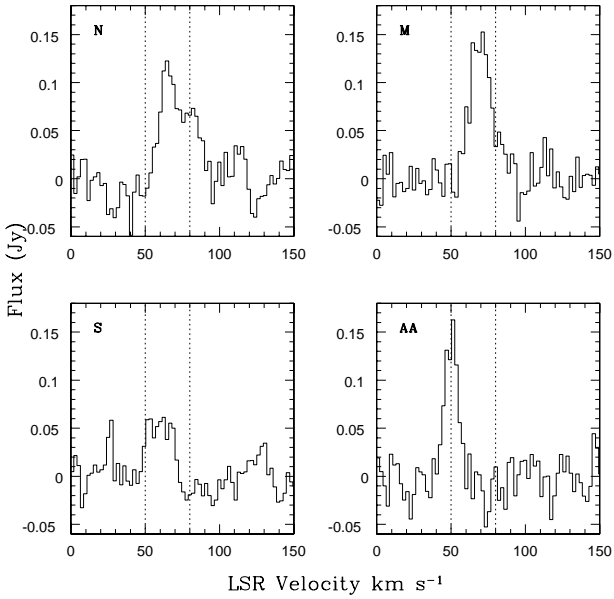,height=3.3truein,width=3.3truein}
\end{center}
\vskip -0.1 in
\caption{ GMRT \acet~ emission spectra towards the four regions marked
in Fig.~\ref{fig:uv10ov}. The dotted lines are at LSR velocities of 50 \kms~
and 80 \kms. The observed continuum flux densities from the regions over which the
spectra have been extracted are 3.4 Jy (N), 4.2 Jy (M), 1.4 Jy (S) and
4.4 Jy (AA). The peak line to continuum ratio is thus $\lesssim 0.035$ in
all cases. See text for discussion.}
\label{fig:uv10spc}
\end{figure}

	\acet~emission was detected from Sgr~B2 over the LSR velocity range
$45 - 85$~\kms. Fig.~\ref{fig:uv10ov} shows the integrated emission from
this entire velocity range (contours) overlaid on the continuum emission
at 1065~MHz (grey scale).  We detect emission from the regions \sgr~ north 
(N), main (M) and south (S), as well as from a location $\sim 4'$ to the 
south of \sgr~(N) (which is labelled AA in Fig.~\ref{fig:uv10ov} after
Mehringer et al. 1993). The spectra of  these regions (Fig.~\ref{fig:uv10spc})
show a systematic velocity gradient in the north-south direction, with 
\sgr~(N) showing emission up to an LSR velocity of $\sim 85$~\kms~and 
\sgr~(AA) having emission centered at $\sim 50$~\kms. A similar gradient 
is seen in several molecular species associated with \sgr, e.g. HC$_{\rm 3}$N 
and CH$_{\rm 3}$CN (\cite{devicente97}) and H$_2$CO (\cite{rogstad74}, 
\cite{mp90}, \cite{mehringer95}). We do not detect any \acet~emission 
towards Sgr~B1, although it lies within our field of view and our observing 
frequency band is sufficiently wide to include any emission associated with it.

	The other noteworthy feature of Fig.~\ref{fig:uv10ov} is that the 
line emission is confined to regions with continuum emission. This implies that
either the \acet~emission is mased or that the gas which produces the 
\acet~emission is closely associated with that producing the radio continuum.
However, the velocities of the \acet~ emission match those of the formaldehyde
(H$_2$CO) absorption in the different regions (\cite{mp90,mehringer95}), showing
that the \acet~ emitting clouds lie in front of the HII regions which are the
source of the radio continuum. In the absence of masing, this would require 
fine-tuning of physical conditions to ensure the excitation of gas only in 
clouds in front of the continuum sources (the parent gas cloud, as traced by
the CH$_{\rm 3}$CN and HC$_{\rm 3}$N emission,  is more extended than the 
region showing \acet~emission; \cite{devicente97}). The observed line to 
continuum ratio in all four regions is $\sim 0.035$, consistent with the 
\acet~ emission being weakly mased (the continuum flux densities of the regions 
over which the spectra in Fig.~\ref{fig:uv10spc} were extracted are listed in the
figure caption). Note, however, that the continuum emission in \sgr~ is 
considerably more extended than the line emission (and hence more resolved out 
in our interferometric observations); the measured line to continuum ratios
in our maps are thus {\it upper limits} on the true line to continuum ratios. 
Our conclusion  that the observed emission is mased is consistent with that of 
Fourikis et al. (1974), who found that the relative intensities of the $1_{10} 
\rightarrow 1_{11}$  and $2_{11} \rightarrow  2_{12}$ lines could only be 
explained if the former line is weakly mased. 

\begin{figure}[t!]
\begin{center}
\epsfig{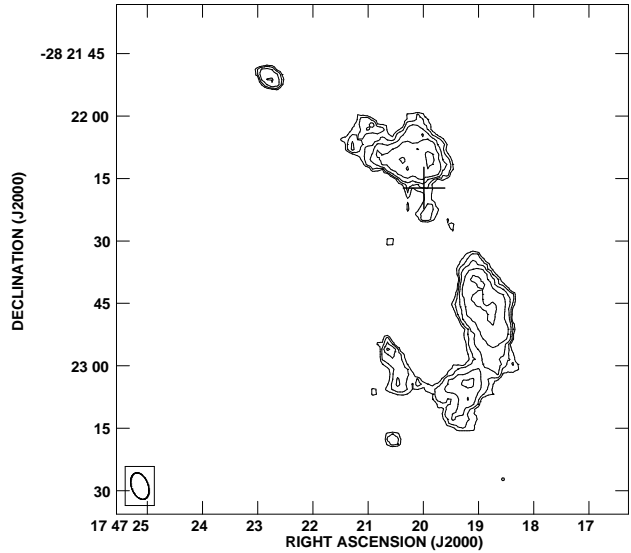}
\end{center}
\vskip -0.1 in
\caption{ Integrated \acet~ emission towards \sgr. The spatial 
resolution is $6.6'' \times 3.8''$. The contour levels are at 30, 48.8, 79.4,
129.1 and 210~Jy/Beam~m/s. The position of \sgr~(LMH) is marked with a cross 
on the figure. 
}
\label{fig:uv60m0}
\end{figure}

        Fig.~\ref{fig:uv60m0} is a high resolution ($6.6''\times3.8''$) 
image of the velocity integrated \acet~ emission.  \sgr~(AA) is not detected 
at this high resolution. The emission correlates very well with the high resolution
images of 4.8~GHz formaldehyde absorption (\cite{whiteoak83}), the 6.7~GHz methanol 
absorption (\cite{houghton95}) and the quasi-thermal 44~GHz methanol emission 
(\cite{mehringer97b}). It also overlaps with the star forming ridge recently
detected via emission from vibrationally excited HC$_3$N in 
\sgr~(\cite{devicente00}). 

	Fig.~\ref{fig:uv10ov} shows clearly that, unlike most other large organic 
molecules, \acet~ is wide-spread in \sgr. The angular extent of the emission
is, in fact, reasonably close to that assumed by Fourikis et al. (1974) in
their computation of the excitation temperature and column density of
\acet. If we assume that the $2_{11} \rightarrow  2_{12}$ transition of
\acet~ observed by Fourikis et al. (1974) is thermalized and has an excitation 
temperature of $\sim 100$~K, we can use their eqn.~(3) to obtain a \acet~
column density of $\sim 10^{14}$~cm$^{-2}$ in the $2_{12}$ level. On the other 
hand, the column density in the 1$_{11}$ level of H$_2$CO is $\sim 10^{14} 
\times T_{\rm ex}$ cm$^{-2}$ (\cite{mp90}). For reasonable values of the 
H$_2$CO $1_{10}~ \rightarrow~1_{11}$ excitation temperature (i.e. few K), and 
under the further assumption that the abundance ratio obtained from these two 
levels is representative of the actual abundance ratio of the molecules, 
the ratio of the molecular column densities is N(H$_2$CO)/N(\acet) $\sim 1$. 
This is considerably smaller than the ratio of $\sim 10^{4}$ expected from pure
gas phase models for the production of these molecules (\cite{lee96}) and suggests
that grain chemistry is probably important for the production of \acet.  

	The  enhanced wide-spread abundance of \acet~ is probably related to 
the presence of numerous shocks in \sgr, for which there are several independent 
lines of evidence. (1) Sato et al. (2000) postulate a large-scale shock resulting 
from a cloud-cloud collision. Evidence for such a collision comes from large-scale CO 
maps as well as the presence of various molecular line masers (e.g. CH$_3$OH, 
H$_2$CO; \cite{houghton95,whiteoak83}) and distorted magnetic fields 
(\cite{dowell98}), all of which are aligned approximately north-south, 
broadly coincident with the \acet~emission of Fig.~\ref{fig:uv10ov}. Further, 
the eastern edge of the northern section and the western edge of the southern 
section of the \acet~emission seen in Fig.~\ref{fig:uv10ov} lie along the 
boundary of this postulated cloud-cloud collission.  (2) Mehringer \& Menten (1997)
argue that the quasi-thermal 44~GHz methanol emission (which is spatially 
correlated with the \acet~emission) probably arises from shocked gas 
at the boundaries of expanding ionized shells around the young stars in 
the molecular cloud. (3) Mart\'{i}n-Pintado et al. (1999) have found a large number 
of hot expanding shells produced by massive evolved stars in the region between 
\sgr~(M) and \sgr~(AA), from which we detect bright \acet~ emission. Again, 
the expansion of these hot shells into the parent molecular cloud is likely 
to result in the formation of shocks. Thus, shocks do appear to be present 
at a number of locations in \sgr. These shocks can cause the disruption of the 
grain mantles on which organic molecules have been formed and thus result in 
the release of these molecules into the gas phase. This could account for the 
observed wide-spread emission from \acet.

In time-dependent astro-chemical models, the abundance of complex 
molecules released from grain mantles rapidly decreases with time, due to 
their destruction via UV photons or collisions (e.g. \cite{charnley92,charnley95}).
The non-detection of acetaldehyde towards Sgr~B1 (which is believed to be 
considerably older than \sgr; \cite{mehringer92}) is thus consistent with models 
in which this molecule is produced by the (possibly shock-induced) disruption of 
grain mantles.  We note, however, that the mechanism for inverting the populations 
of the \acet~$1_{11}$ and $1_{10}$ levels is currently unknown; it is thus also 
possible that \acet~ is present in Sgr~B1, but that physical conditions here 
are not appropriate to excite the 1065~MHz transition. New interferometers, such 
as the GMRT, which can yield wide field images of molecular emission,
now make it possible to obtain a clearer view of the large-scale distribution 
of complex organic molecules in the Galaxy.

\begin{acknowledgements}
	These observations would not have been possible without the many years 
of dedicated effort put in by the GMRT staff in order to build the telescope.
The GMRT is operated by the National Centre for Radio Astrophysics of the Tata
Institute of Fundamental Research. We thank the referee, J. Mart\'{i}n-Pintado,
for his comments and suggestions which have improved this paper.
\end{acknowledgements}

\end{document}